\documentclass[a4paper,10pt]{article}
\usepackage[dvips]{graphicx}
\usepackage{amssymb,amsmath}
\usepackage{upgreek}
\usepackage{color}

\numberwithin{equation}{section}

\begin{document}

 \def\gsim{ \lower .75ex \hbox{$\sim$} \llap{\raise .27ex \hbox{$>$}} }
 \def\lsim{ \lower .75ex \hbox{$\sim$} \llap{\raise .27ex \hbox{$<$}} }



\title{\bf Integrable (2 + 1)-Dimensional Spin Models with \\ Self-Consistent Potentials}

\author{Ratbay Myrzakulov$^1$\footnote{Email: rmyrzakulov@gmail.com}, Galya Mamyrbekova$^1$\footnote{Email: galiya110160@mail.ru}, Gulgassyl Nugmanova$^1$\footnote{Email: nugmanovagn@gmail.com} \\ and Muthusamy Lakshmanan$^2$\footnote{Email: lakshman@cnld.bdu.ac.in}
\\ \textit{$^1$ Eurasian International Center for Theoretical Physics and  Department of General,}  \\\textit{ Theoretical Physics, Eurasian National University, Astana 010008, Kazakhstan}\\
\textit{$^2$ Centre for Nonlinear Dynamics, School of Physics, }\\ \textit{Bharathidasan University, Tiruchirappalli 620 024, India } }

\date{}
 \maketitle


 \renewcommand{\baselinestretch}{1.1}

{\bf Abstract}: Integrable spin systems possess interesting geometrical and gauge invariance properties and have important applications in applied magnetism and nanophysics. They are also intimately connected to the nonlinear Schr\"odinger family of equations. In this paper, we identify three different integrable spin systems in (2 + 1) dimensions by introducing the interaction of the spin field with more than one scalar potential, or vector potential, or both. We also obtain the associated Lax pairs. We discuss various interesting reductions in (2 + 1) and (1 + 1) dimensions. We also deduce the equivalent nonlinear Schr\"odinger family of equations, including the (2 + 1)-dimensional version of nonlinear Schr\"odinger--Hirota--Maxwell--Bloch equations, along with their Lax pairs.

\tableofcontents
\section{Introduction}
Integrable and non-integrable spin systems \cite{royal} play a very useful role in nonlinear \linebreak physics and mathematics. They give rise to important applications in applied magnetism \cite{hillebrands} and nanophysics \cite{bertotti}. The Landau--Lifshitz--Gilbert (LLG) equation \cite{stiles} in ferromagnetism and Landau--Lifshitz--Gilbert--Slonczewski (LLGS) equation \cite{bertotti} in spin transfer nanomagnetic multilayers are some of the fundamental equations that play a crucial role in understanding various physical properties of magnetic materials and the development of new technological innovations, like microwave generation using the spin transfer effect \cite{slonczewski}. The continuum limit of the Heisenberg ferromagnetic spin system and its various generalizations give rise to some of the important integrable spin systems in (1 + 1) dimensions \cite{lakshmanan77,takhtajan}. They are also intimately related to the nonlinear Schr\"odinger family of equations through geometrical (or Lakshmanan equivalence or L-equivalence) and gauge equivalence concepts and these systems often admit magnetic soliton solutions \cite{royal}.

Though a straightforward generalization of the (1 + 1)-dimensional Heisenberg spin system to \linebreak (2 + 1) dimensions is not integrable \cite{senthilkumar}, inclusion of additional terms corresponding to the interaction of a scalar potential field makes the spin system integrable. The well-known Ishimori equation \cite{ishimori} and the Myrzakulov I equation \cite{myrzakulov-391} are two of the most interesting integrable spin equations. Their geometrical and gauge-equivalent counterparts are the Davey--Stewartson and Zakharov--Strachan equations \cite{ablowitz}, respectively. They admit (2 + 1)-dimensional localized structures \cite{ablowitz,myrzakulov-2122}. Interestingly, such an interaction of the spin vector with the scalar potential can be further generalized. One can include more than one scalar potential and make them interact with the spin vector to generate new integrable spin equations. Furthermore, one can even introduce the interaction of a vector (unit) potential with the spin vector. The result is that one can obtain more general integrable (2 + 1)-dimensional spin evolution equations along with their associated Lax pairs. In this paper, we introduce three such integrable spin models in (2 + 1) dimensions, namely Myrzakulov--Lakshmanan (ML) II, III and IV equations. We also point out that equivalent (2 + 1)-dimensional integrable nonlinear Schr\"odinger--Maxwell--Bloch-type evolution equations and their Lax pairs can also be identified. From these equations, several interesting limiting cases of nonlinear evolution equations in (2 + 1) and (1 + 1) dimensions, along with their Lax pairs, can also be deduced. In this paper, we do not attempt to explicitly solve the initial value problem associated with the Lax pair and obtain explicit localized solutions, which will be reported in \linebreak subsequent works.

The plan of the paper is as follows.
In Section 2, we give the basic facts from the theory of the generalization of the Heisenberg ferromagnetic spin equation in (2 + 1) dimensions. 
In Section 3, we investigate the ML-II equation. 
Next, we study the ML-III equation in Section 4. In Section 5, 
we consider the ML-IV equation. 
Finally, we give our conclusions in Section 6.  

\section{A Brief Review on Integrable Spin Systems in 2 + 1 Dimensions}

There exists a few integrable spin systems in (2 + 1) dimensions in the literature \cite{ishimori,myrzakulov-391,ablowitz,myrzakulov-2122,myrzakulov-3765}. In this section, we present some basic features of them.

\subsection{The Ishimori equation}
The well-known Ishimori equation has the form \cite{ishimori}
\begin{eqnarray}
{\bf S}_{t}-{\bf S}\wedge ({\bf S}_{xx}+{\bf S}_{yy})-u_x{\bf S}_{y}-u_y{\bf S}_{x}&=&0\label{2.1},\\
 u_{xx}-\upalpha^2u_{yy}+2\upalpha^2{\bf S}\times({\bf S}_{x}\wedge {\bf S}_{y})&=&0\label{2.2},
 \end{eqnarray}
where $\wedge$ denotes the vector product (or the cross product); $\upalpha=const$; ${\bf S}$ is the spin vector with unit length, that is: \begin{equation}\label{2.3}
{\bf S}=(S_1,S_2,S_3), \quad {\bf S}^2=1,
\end{equation}
and $u$ is a scalar real function (potential). The Ishimori equation admits the following Lax \linebreak representation \cite{slonczewski},
\begin{eqnarray}
\Phi_{x}+\upalpha S\Phi_y&=&0\label{2.4},\\
\Phi_{t}-A_2\Phi_{xx}-A_1\Phi_{x}&=&0\label{2.5},
\end{eqnarray} 
where 
 \begin{eqnarray}
A_2&=&-2i S,\label{2.6}\\
A_1&=&-i S_x-i\upalpha S_yS+u_yI-\upalpha^3u_xS\label{2.7}.
\end{eqnarray} 
here
\begin{equation}\label{2.8}
S=S_i\upsigma_i=\begin{pmatrix} S_3 &S^{-}\\S^{+} & -S_3\end{pmatrix},
 \end{equation}
where
$S^2=I, \quad S^{\pm}=S_1\pm iS_2, \quad [A,B]=AB-BA,\quad I=diag(1,1)$ is the identity matrix and the $\upsigma_i$ are the Pauli matrices,
\begin{equation}\label{2.9}
\upsigma_1=\begin{pmatrix} 0&1\\1& 0\end{pmatrix}, \quad \upsigma_2=\begin{pmatrix} 0&-i\\i& 0\end{pmatrix}, \quad \upsigma_3=\begin{pmatrix} 1&0\\0& -1\end{pmatrix}.
 \end{equation}
 The Ishimori equation is one of the integrable (2 + 1)-dimensional extensions of the following celebrated integrable (1 + 1)-dimensional continuum Heisenberg ferromagnetic spin equation (HFE) \cite{lakshmanan77,takhtajan},
 \begin{equation}\label{2.10}
{\bf S}_{t}-{\bf S}\wedge {\bf S}_{xx}=0.
\end{equation}
Note that the gauge/geometric equivalent counterpart of the Ishimori equation is the Davey--Stewartson equation \cite{ablowitz}, which reads as:
 \begin{eqnarray}
 i\upvarphi_t + \upalpha^2\upvarphi_{xx} +\upvarphi_{yy} -v\upvarphi+2|\upvarphi|^2\upvarphi&=&0\label{2.11},\\
 v_{xx}-\upalpha^2v_{yy}+4(|\upvarphi|^2)_{yy} &=&0\label{2.12}.
\end{eqnarray}
It is one of the (2 + 1)-dimensional integrable extensions of the nonlinear Schr\"odinger equation (NSE)
 \begin{equation}
i\upvarphi_{t}+\upvarphi_{xx}+2|\upvarphi|^2\upvarphi=0\label{2.13}.
 \end{equation}
Different properties of the Ishimori and Davey--Stewartson equations are well studied in the \linebreak literature \cite{ishimori,ablowitz,myrzakulov-2122}. Furthermore, we can recall that between the HFE (\ref{2.10}) and NSE (\ref{2.13}), the Lakshmanan and gauge equivalence takes place \cite{royal,lakshmanan77,takhtajan}. 

{Both Ishimori and D-S equations admit Lax pairs and, so, are linearizable. They admit a zero-curvature representation and, so, are integrable in the Lax sense. } {The Riemann--Hilbert problems associated with the linear eigenvalue problems have been analyzed \cite{K2, Fokas} to solve the initial value problems for appropriate boundary conditions.} { The resulting solutions in the form of solitons, exponentially-localized dromions, \emph{etc.}, have been given \cite{K1}, and Hirota bilinearization has been effected.} {Furthermore, the associated infinite number of involutive integrals of motion has been obtained.} {In this sense, both Ishimori and the D-S equation} {are} {considered to be integrable (2 + 1)-dimensional nonlinear evolution equations} \cite{ablowitz, K1}.

\subsection{The Myrzakulov-I  equation}
As the second example of the integrable spin systems in (2 + 1) dimensions, we here present some details of the Myrzakulov-I  equation  (M-I equation) \cite{myrzakulov-3765}. It reads as
\begin{eqnarray}
{\bf S}_{t}-{\bf S}\wedge {\bf S}_{xy}-u{\bf S}_{x}&=&0 \label{2.14},\\
 u_x+{\bf S}\times({\bf S}_{x}\wedge {\bf S}_{y})&=&0\label{2.15}. \end{eqnarray} 
 Often, we write this equation in the following form
 \begin{eqnarray}
{\bf S}_{t}-({\bf S}\wedge {\bf S}_{y}+u{\bf S})_{x}&=&0\label{2.16},\\
 u_x+{\bf S}\times({\bf S}_{x}\wedge {\bf S}_{y})&=&0\label{2.17}. \end{eqnarray} 
 The M-I equation has the following Lax representation
\begin{eqnarray}
\Phi_{x}-\frac{i}{2}\uplambda S\Phi&=&0\label{2.18},\\
\Phi_{t}-\uplambda\Phi_{y}-\uplambda Z\Phi&=&0\label{2.19}, 
\end{eqnarray} 
where
 \begin{equation}
 Z=\frac{1}{4}([S,S_y]+2iuS)\label{2.20}.
\end{equation} 
Note that for this equation, the eigenvalue satisfies the equation
\begin{equation}\label{lam}
 \uplambda_{t}=2\uplambda\uplambda_y.
\end{equation} 

Some properties of the M-I equation were studied, for example, in \cite{myrzakulov-391,myrzakulov-2122,Chen,Hai}.
Like the Ishimori equation, the M-I equation is one of the (2 + 1)-dimensional extensions of the (1 + 1)-dimensional HFE (\ref{2.10}). The gauge/geometric equivalent counterpart of the M-I equation has the form \cite{myrzakulov-3765}
\begin{eqnarray}
 i\upvarphi_t + \upvarphi_{xy} -v\upvarphi&=&0\label{2.21},\\
 v_{x}+2(|\upvarphi|^2)_{y}&=&0\label{2.22},
 \end{eqnarray}
which is nothing but one of the (2 + 1)-dimensional extensions of the NSE (\ref{2.13}) \cite{myrzakulov-2122}. This equation is known as the Zakharov--Strachan equation (see \cite{Zakharov,Strachan1,Strachan2}).

Additionally, here, we can note that Lax representations involving linear problems where the eigenvalue evolves as {a function of time (and even as a function of x) are already well known} \cite{Calogero1, Calogero2, LB, Bala1, L31,  Bala2, Bala3} in {the case} {of the (1 +1)-dimensional Heisenberg ferromagnetic spin evolution equation with linearly x}-{dependent terms, the gauge-equivalent nonlinear Schr\"odinger equation with linearly x-dependent terms and their inverse scattering analysis, and} {an infinite number of conserved quantities has been established.} {However, we note here that in the case of the present (2 + 1)-dimensional problem, for Equation (\ref{lam}), $\uplambda (t) = \uplambda_0 = const$ is also a solution, so that the associated} {Riemann--Hilbert problem is similar to the already studied integrable cases.} {However, the existence of a more general evolution Equation (\ref{lam})} {for the eigenvalue implies} {that an even richer structure of solutions exists, which remains to be fully explored.}

 \subsection{The Myrzakulov--Lakshmanan I Equation}
 
 Another example of the integrable spin systems in (2 + 1) dimensions is the so-called Myrzakulov--Lakshmanan I  equation (ML-I equation)\cite{Es}, which reads as
 \begin{eqnarray}
{\bf S}_{t}-{\bf S}\wedge (\upalpha S_{xx}+\upbeta{\bf S}_{xy})-u{\bf S}_{x}&=&0\label{2.23},\\
 u_x+{\bf S}\times({\bf S}_{x}\wedge {\bf S}_{y})&=&0\label{2.24}. \end{eqnarray} 
 It has the following Lax representation
\begin{eqnarray}
\Phi_{x}-\frac{i}{2}\uplambda S\Phi&=&0\label{2.25},\\
\Phi_{t}-\uplambda\upbeta\Phi_{y}-B\Phi&=&0\label{2.26}, 
\end{eqnarray} where 
 \begin{equation}
B=\upalpha(\frac{1}{2}i\uplambda^2 S+\frac{1}{4}[S,S_x])+\upbeta\uplambda Z\label{2.27}.
\end{equation} 
 The Myrzakulov--Lakshmanan I equation is another integrable (2 + 1)-dimensional extension of the \linebreak (1 + 1)-dimensional Heisenberg ferromagnet equation (\ref{2.10}). {The ML-I equation admits the well-known two integrable reductions: the HFE (\ref{2.10}) as $\upalpha=1, \quad \upbeta=0$ and the M-I Equations (\ref{2.14}) and (\ref{2.15}) as $\upalpha=0, \quad \upbeta=1$.} {As for its reductions, HFE and M-I equations, for the ML-I equation, we can also mention the existence of Hirota bilinearization, soliton and dromion solutions and other ingredients of integrable systems (see \cite{myrzakulov-391, myrzakulov-2122, myrzakulov-3765}).} {Finally, note that the equivalent counterpart of the Myrzakulov--Lakshmanan I equation is the evolution equation }
 \begin{eqnarray}
 i\upvarphi_t + \upalpha\upvarphi_{xx} +\upbeta\upvarphi_{xy} +v\upvarphi&=&0\label{2.28},\\
 v_{x}-2[\upalpha(|\upvarphi|^2)_{x}+\upbeta(|\upvarphi|^2)_{y}] &=&0\label{2.29},
\end{eqnarray}
{which is also one of the integrable (in the Lax representation sense) (2 + 1)-dimensional extensions of the nonlinear Schr\"odinger equation (\ref{2.13}).} {This equation admits two integrable reductions: the famous NSE (\ref{2.13}) as $\upalpha=1, \quad \upbeta=0$ and the Zakharov--Strachan Equations (\ref{2.22}) and (\ref{2.23}) as $\upalpha=0, \quad \upbeta=1$.} {The Lax representation of the Equations (\ref{2.28}) and (\ref{2.29}) is given by
\begin{eqnarray}
\Psi_{x}&=&U_2\Psi \label{2.31cc},\\
\Psi_{t}&=&\uplambda\upbeta\Psi_{y} +V_2\Psi \label{2.31c},
\end{eqnarray}
where
\begin{eqnarray}
U_2=\frac{i\uplambda}{2}\upsigma_3+G, \quad G=\left(
\begin{array}{cc}
0 & \upvarphi^{*} \\
\upvarphi & 0
\end{array}\right), \quad V_2=\frac{i\uplambda^2}{2}\upalpha\upsigma_3+\upalpha\uplambda G+V_0 \label{2.33c}.
\end{eqnarray}
here
\begin{eqnarray}
 V_0=\left(
\begin{array}{cc}
\upalpha i|\upvarphi|^{2}+i\upbeta\partial_x^{-1}|\upvarphi|^{2}_y & -i\upbeta \upvarphi^{*}_y-i\upalpha \upvarphi^{*}_x \\
i\upbeta \upvarphi_y+\upalpha i\upvarphi_x & -[\upalpha i|\upvarphi|^{2}+i\upbeta\partial_x^{-1}|\upvarphi|^{2}_y]
\end{array} \right) \label{2.34c}.
\end{eqnarray}
Note that in this case, the spectral parameter $\uplambda$ obeys the equations:
\begin{eqnarray}
\uplambda_t=\upbeta\uplambda\uplambda_y, \quad \uplambda_x=0 \label{2.35c}
\end{eqnarray}
that is is the function of $y$ and $t$.}

 \subsection{The (2 + 1)-Dimensional Heisenberg Ferromagnet Equation}
  The physically-important (2 + 1)-dimensional Heisenberg ferromagnet equation can be written as
 \begin{equation}
{\bf S}_{t}={\bf S}\wedge ({\bf S}_{xx}+{\bf S}_{yy}) \label{2.30}.
\end{equation}
It is a very important system from a physical application point of view. However, unfortunately, this equation is not integrable \cite{senthilkumar}.
\section{The Myrzakulov-Lakshmanan II equation}

In this and the next two sections, we will present a new class of integrable spin systems in (2 + 1) dimensions by introducing a vector potential interacting with the spin field self-consistently in addition to the scalar potential considered in Section 2. We will give their Lax representations, equivalent counterparts and some reductions. We start from the so-called Myrzakulov--Lakshmanan II equation (ML-II equation), which has the form
\begin{eqnarray}
{\bf S}_{t}-{\bf S}\wedge{\bf S}_{xy}-u{\bf S}_{x}-\frac{1}{\upomega}{\bf S}\wedge {\bf W}&=&0\label{3.1},\\
u_x+{\bf S}\times(\bf S}_{x}\wedge{\bf S_{y})&=&0\label{3.2},\\
 {\bf W}_{x}-\upomega {\bf S}\wedge{\bf W}&=&0\label{3.3}
\end{eqnarray} 
or equivalently
\begin{eqnarray}
iS_{t}+\frac{1}{2}[S, S_{xy}]+iuS_{x}+\frac{1}{\upomega}[S, W]&=&0\label{3.4},\\
u_x-\frac{i}{4}tr(S[S_x,S_y])&=&0\label{3.5},\\
 iW_{x}+\upomega [S, W]&=&0\label{3.6}.
\end{eqnarray} 
here, $S=S_i\upsigma_i$, $W=W_i\upsigma_i$, $i=1,2,3$ and $\upomega$ is a constant parameter. The vector ${\bf W}=(W_1,W_2,W_3)$ may be considered as a vector potential.
 The ML-II equation is linearizable, possesses a Lax pair and, so, is expected to be integrable like the previous equations. In the next subsections, we give some important information on this equation. 
\subsection{Reductions} 

Some comments on the reduction of the ML-II equation are in order. First, we note that if we put $W=0$, then the ML-II Equations (\ref{3.1})--(\ref{3.3}) reduce to the M-I Equations (\ref{2.14}) and (\ref{2.15}). If we consider the case $y=x$, then the ML-II Equations (\ref{3.1})--(\ref{3.3}) transform to the following Myrzakulov-XCIX equation (M-XCIX equation) (see e.g. \cite{R13} and \cite{ML2014}),
 \begin{eqnarray}
iS_{t}+\frac{1}{2}[S, S_{xx}]+\frac{1}{\upomega}[S, W]&=&0\label{3.7},\\
 iW_{x}+\upomega [S, W]&=&0\label{3.8}.
\end{eqnarray} 
 Therefore, the ML-II equation is one of the potential (2 + 1)-dimensional integrable extensions of the M-XCIX equation. 
\subsection{Lax Representation}

 The ML-II Equations (\ref{3.1})--(\ref{3.3}) are integrable in the sense that they can be associated with a linear
 eigenvalue problem and that they admit a Lax representation. The corresponding Lax representation can be written in the form
 \begin{eqnarray}
\Phi_{x}&=&U\Phi\label{3.9},\\
\Phi_{t}&=&2\uplambda\Phi_y+V\Phi\label{3.10}. 
\end{eqnarray} 
here, the matrix operators $U$ and $V$ have the forms 
 \begin{eqnarray}
U&=&-i\uplambda S\label{3.11},\\
V&=&\uplambda V_{1}+\frac{i}{\uplambda+\upomega}W-\frac{i}{\upomega}W\label{3.12},
\end{eqnarray} 
where
\begin{eqnarray}
V_1&=&2Z=\frac{1}{2}([S, S_{y}]+2iuS)\label{3.13},\\
W&=&\begin{pmatrix} W_3&W^{-}\\W^{+}& -W_3\end{pmatrix}\label{3.14}. 
\end{eqnarray} 

{We also note that here, also, the spectral parameter obeys the equation $\uplambda_t=2\uplambda\uplambda_y,$} \textcolor{black}{that is, we can associate a nonisospectral problem.} \textcolor{black}{As noted above, any constant solution of this equation \linebreak $\uplambda=\uplambda_{0}=constant$ } \textcolor{black}{should act as an eigenvalue parameter for the corresponding Lax representation. However, its non-constant solutions can also serve as the spectral parameter and can be bring out the rich nature of solutions.} \textcolor{black}{For example, this equation for the spectral parameter has the following particular solution: $\uplambda=(\upbeta_1+\upbeta_2y)(\upbeta_3-2\upbeta_1t)^{-1}$, where $\upbeta_j$, j=1,2,3 are } \textcolor{black}{in general some complex constants. It is interesting to note that this variable form of the spectral parameter} \textcolor{black}{gives us the physically-interesting solution, the so-called shock wave solution of the original nonlinear evolution equation.} \textcolor{black}{Here, we can note that the above presented} \textcolor{black}{equation is in fact the well-known Riemann equation for shock waves. Lastly, we wish to recall the fact that the occurrence of} \textcolor{black}{the variable spectral parameters or, in other words, nonisospectral parameters (that is, the $\uplambda$ is some function of $(t), \quad (t,x), \quad (t, y)$ or $(t,x, y)$) have been} \textcolor{black}{well appreciated in soliton theory from long ago (see the seminal papers \cite{Calogero2, LB, Calogero3, Calogero4}).} \textcolor{black}{Using such a nonisospectral parameter approach,} \textcolor{black}{recently, several physically-interesting
nonlinear evolution equations with external potentials were constructed by Sakhnovich \cite{Sa}. }

 \subsection{Gauge Equivalent Counterpart of the ML-II Equation}

 Let us find the gauge equivalent counterpart of the ML-II Equations (\ref{3.1})--(\ref{3.3}). It is not difficult (see Appendix A) to verify that the gauge-equivalent counterpart of the ML-II equation 
 is given by
\begin{eqnarray}
q_{t}+\frac{\kappa}{2i}q_{xy}+ivq-2p&=&0 \label{3.15},\\
r_{t}-\frac{\kappa}{2i}r_{xy}-ivr-2k&=&0\label{3.16},\\
v_{x}+\frac{\kappa}{2}(rq)_{y}&=&0\label{3.17},\\
p_{x}-2i\upomega p -2\upeta q&=&0\label{3.18},\\
k_x+2i\upomega k-2\upeta r&=&0\label{3.19},\\
\upeta_{x}+r p +k q&=&0\label{3.20},
 \end{eqnarray}
where $q,r,p, k$ are some complex functions; $v, \upeta$ are potential functions and $\kappa$ is a constant parameter. We call this set of equations the (2 + 1)-dimensional nonlinear Schr\"odinger--Maxwell--Bloch equation (NSMBE) due to the fact that in 1 + 1 
 dimensions, it reduces to the (1 + 1)-dimensional nonlinear Schr\"odinger--Maxwell--Bloch equation (see, e.g., \cite{C1,C2} and the references therein). 
Of course, this equation is also integrable in the Lax sense. The corresponding Lax representation to \linebreak Equations (\ref{3.15})--(\ref{3.20}) reads as
\begin{eqnarray}
\Psi_{x}&=&A\Psi\label{3.21},\\
\Psi_{t}&=&\kappa\uplambda\Psi_y+B\Psi\label{3.22}, 
\end{eqnarray} 
where \textcolor{black}{the spectral parameter $\uplambda$ obeys the evolution equation
$\uplambda_{t}=\kappa\uplambda\uplambda_{y}$ and}
 \begin{eqnarray}
A&=&-i\uplambda \upsigma_3+A_0\label{3.23},\\
B&=&B_0+\frac{i}{\uplambda+\upomega}B_{-1}\label{3.24}. 
\end{eqnarray} 
here
\begin{eqnarray}
A_0&=&\begin{pmatrix} 0&q\\-r& 0\end{pmatrix}\label{3.25},\\
B_0&=&-\frac{i}{2}v\upsigma_3-\frac{\kappa}{2i}\begin{pmatrix} 0&q_y\\r_y& 0\end{pmatrix}\label{3.26},\\
B_{-1}&=&\begin{pmatrix} \upeta&-p\\-k& -\upeta\end{pmatrix}\label{3.27}. 
\end{eqnarray}

Next, we consider the reduction $r=\updelta q^{*}, \quad k=\updelta p^{*}$ with $\kappa=2$, where $*$ means the complex conjugate. Then, System (\ref{3.15})--(\ref{3.20}) takes the form
 \begin{eqnarray}
iq_{t}+q_{xy}-vq-2ip&=&0 \label{3.45},\\
v_{x}+2\updelta(|q|^2)_{y}&=&0\label{3.46},\\
p_{x}-2i\upomega p -2\upeta q&=&0\label{3.47},\\
\upeta_{x}+\updelta(q^{*} p +p^{*} q)&=&0 \label{3.48},
\end{eqnarray}
where we have assumed that $\updelta=\pm 1$. We note that in (1 + 1) dimensions, that is if $y=x$, the last system takes the form
\begin{eqnarray}
iq_{t}+q_{xx}+2\updelta |q|^2q-2ip&=&0 \label{3.49},\\
p_{x}-2i\upomega p -2\upeta q&=&0\label{3.50},\\
\upeta_{x}+\updelta(q^{*} p +p^{*} q)&=&0\label{3.51},
\end{eqnarray}
which is nothing but the well-known (1 + 1)-dimensional nonlinear Schr\"odinger--Maxwell--Bloch equation (see, e.g., \cite{C1,C2} and the references therein). Its Lax pair has the form
\begin{eqnarray}
\Psi_{x}&=&A\Psi\label{3.52},\\
\Psi_{t}&=&2\uplambda A\Psi+B\Psi\label{3.53},
\end{eqnarray} 
where $A$ and $B$ have the form (\ref{3.23})--(\ref{3.24}) with
\begin{eqnarray}
A_0&=&\begin{pmatrix} 0&q\\-\updelta q^{*}& 0\end{pmatrix}\label{3.54},\\
B_0&=&i\updelta|q|^2\upsigma_3+i\begin{pmatrix} 0&q_x\\\updelta q^{*}_x& 0\end{pmatrix}\label{3.55},\\
B_{-1}&=&\begin{pmatrix} \upeta&-p\\-\updelta p^{*}& -\upeta\end{pmatrix}\label{3.56}.
\end{eqnarray}
Note that the spin-equivalent counterpart of System (\ref{3.49})--(\ref{3.51}) is given by
\begin{eqnarray}
iS_{t}+\frac{1}{2}[S, S_{xx}]+\frac{1}{\upomega}[S, W]&=&0\label{3.57},\\
 iW_{x}+\upomega [S, W]&=&0\label{3.58}.
\end{eqnarray} 
It is nothing but the (1 + 1)-dimensional M-XCIX equations (\ref{3.7}) and (\ref{3.8}), which is well known to be integrable \cite{R13}.
 
 \subsection{Integral of Motion}

 Note that the (2 + 1)-dimensional SMBE (3.25)--(3.27) admits the following integral of motion
 \begin{eqnarray}
I_{1}=\int\int |q|^2dxdy\label{3.43}.
\end{eqnarray} 
In fact, from System (3.25)--(3.27), it follows that
\begin{eqnarray}
(|q|^2)_{t}=-\frac{1}{2}i(q^*_{y}q-q^{*}q_{y}-\frac{2i}{\updelta}\upeta)_{x}-\frac{1}{2}i(q^*_{x}q-q^{*}q_{x})_{y}\label{3.43}.
\end{eqnarray} 
or:
\begin{eqnarray}
(|q|^2)_{t}=div {\bf j}\label{3.43}.
\end{eqnarray} 
Here, the vector ${\bf j}$ is given by
\begin{eqnarray}
{\bf j}\equiv\left(j_{1}, \quad j_{2}\right)=\left(-\frac{1}{2}i(q^*_{y}q-q^{*}q_{y}-\frac{2i}{\updelta}\upeta),\quad -\frac{1}{2}i(q^*_{x}q-q^{*}q_{x})\right)\label{3.43}.
\end{eqnarray}
This result with the gauge equivalence gives us the following integral of motion of the ML-II equation
\begin{eqnarray}
J_{1}=\frac{1}{8}\int\int tr(S_{x}^2)dxdy\label{3.43}.
\end{eqnarray} 
that follows from the relation
\begin{eqnarray}
\frac{1}{8} tr(S_{x}^2)=|q|^2\label{3.43}.
\end{eqnarray}
 \section{The Myrzakulov--Lakshmanan III Equation}

 Now, we want to present another new integrable spin system in 2 + 1 dimensions, namely the so-called the Myrzakulov--Lakshmanan III  equation (ML-III equation), which contains a vector and two scalar potentials. Its form is given as
\begin{eqnarray}
iS_{t}+i\upepsilon_2(S_{xy}+[S_x,Z])_{x}+(wS)_{x}+\frac{1}{\upomega}[S, W]&=&0\label{4.1},\\
u_x-\frac{i}{4}tr(S\times[S_x,S_y])&=&0\label{4.2},\\
w_x-\frac{i}{4}\upepsilon_2[tr(S_x^2)]_y&=&0\label{4.3},\\
iW_{x}+\upomega [S, W]&=&0\label{4.4},
\end{eqnarray} 
where $\upomega$ is a constant parameter and
\begin{eqnarray}\label{4.5}
 Z=\frac{1}{4}([S, S_{y}]+2iuS)\label{4.5}.
 \end{eqnarray} 
Here, $w$ is another scalar potential function.
\subsection{Lax Representation} 

As an integrable equation, the ML-III equation admits a Lax representation. It is given by
 \begin{eqnarray}
\Phi_{x}&=&U\Phi\label{4.6},\\
\Phi_{t}&=&4\upepsilon_2\uplambda^2\Phi_y+V\Phi\label{4.7},
\end{eqnarray} 
where
\begin{eqnarray}
U&=&-i\uplambda S\label{4.8},\\
V&=&4\upepsilon_2\uplambda^2Z+\uplambda V_{1}+\frac{i}{\uplambda+\upomega}W-\frac{i}{\upomega}W\label{4.9} 
\end{eqnarray} 
with
\begin{eqnarray}
V_1&=&wS+i\upepsilon_2(S_{xy}+[S_x,Z])\label{4.10},\\
W&=&\begin{pmatrix} W_3&W^{-}\\W^{+}& -W_3\end{pmatrix}\label{4.11}. 
\end{eqnarray} 

\textcolor{black}{Finally, we note that for the ML-III equation, the spectral parameter satisfies the following nonlinear evolution equation $\uplambda_{t}=4\upepsilon_2\uplambda^2\uplambda_{y}$.}

\subsection{Reductions}

Let us now consider some reductions of the ML-III Equation (\ref{4.1})--(\ref{4.4}). 
\subsubsection{Case I: $\upepsilon_2=0$}

In this case, the ML-III equation reduces to the following principal chiral equation (see, e.g., \cite{Beggs})
\begin{eqnarray}
iS_{t}+\frac{1}{\upomega}[S, W]&=&0\label{4.12},\\
 iW_{x}+\upomega [S, W]&=&0\label{4.13}. \end{eqnarray} 
 It is integrable in the sense that it admits the Lax representation. The corresponding Lax representation follows from Equations (\ref{4.6}) and (\ref{4.7}) as $\upepsilon_2=0$, so that we get
 \begin{eqnarray}
\Phi_{x}&=&U\Phi\label{4.6a},\\
\Phi_{t}&=&V\Phi\label{4.7b}, 
\end{eqnarray} 
where
\begin{eqnarray}
U&=&-i\uplambda S\label{4.8a},\\
V&=&\frac{i}{\uplambda+\upomega}W-\frac{i}{\upomega}W\label{4.9b} 
\end{eqnarray} 
with
\begin{eqnarray}
W&=&\begin{pmatrix} W_3&W^{-}\\W^{+}& -W_3\end{pmatrix}\label{4.11b}. 
\end{eqnarray} 
\subsubsection{Case II: $W=0$}

In this case, we get the following integrable equation
\begin{eqnarray}
iS_{t}+i\upepsilon_2(S_{xy}+[S_x,Z])_{x}+(wS)_{x}&=&0\label{4.14},\\
u_x-\frac{i}{4}tr(S\times[S_x,S_y])&=&0\label{4.15},\\
w_x-\frac{i}{4}\upepsilon_2[tr(S_x^2)]_y&=&0\label{4.16}. \end{eqnarray} 
 
\subsubsection{Case III: $y=x$}

This case corresponds to the Myrzakulov-LXIV equation (M-LXIV equation), which reads as (see, e.g., \cite{R13} and \cite{ML2014})
\begin{eqnarray}
iS_{t}+i\upepsilon_2S_{xxx}+\frac{i}{4}\upepsilon_2(tr(S_x^2)S)_{x}+\frac{1}{\upomega}[S, W]&=&0\label{4.17},\\
 iW_{x}+\upomega [S, W]&=&0\label{4.18}.
\end{eqnarray}

\subsection{Equivalent Counterpart of the ML-III Equation}

 It is not difficult to verify that the gauge-equivalent counterpart of the ML-III equation has the form
 \begin{eqnarray}
iq_{t}+i\upepsilon_2q_{xxy}-vq+(wq)_x-2ip&=&0 \label{4.19},\\
ir_{t}+i\upepsilon_2r_{xxy}+vr+(wr)_x-2ik&=&0\label{4.20},\\
v_{x}-2i\upepsilon_2(r_{xy}q-rq_{xy})&=&0\label{4.21},\\
w_{x}-2i\upepsilon_2(rq)_y&=&0\label{4.22},\\
p_{x}-2i\upomega p -2\upeta q&=&0\label{4.23},\\
k_x+2i\upomega k-2\upeta r&=&0\label{4.24},\\
\upeta_{x}+r p +k q&=&0\label{4.25}.
 \end{eqnarray}
 
This equation can be considered as the general (2 + 1)-dimensional complex modified Korteweg--de Vries--Maxwell--Bloch equation (cmKdVMBE), as it is one of the (2 + 1)-dimensional generalizations of the (1 + 1)-dimensional cmKdVMB equation (see, e.g., \cite{C1,C2}).
Of course, Equations (\ref{4.19})--(\ref{4.25}) are also integrable in the Lax sense and due to the gauge equivalence with System (\ref{4.1})--(\ref{4.5}). The corresponding Lax representation reads as
\begin{eqnarray}
\Psi_{x}&=&A\Psi\label{4.26},\\
\Psi_{t}&=&4\upepsilon_2\uplambda^2\Psi_y+B\Psi\label{4.27},
\end{eqnarray} 
where 
 \begin{eqnarray}
A&=&-i\uplambda \upsigma_3+A_0\label{4.28},\\
B&=&\uplambda B_1+B_0+\frac{i}{\uplambda+\upomega}B_{-1}\label{4.29} 
\end{eqnarray} 
\textcolor{black}{and the spectral parameter satisfies the following equation $\uplambda_{t}=4\upepsilon_2\uplambda^2\uplambda_{y}$, which has the same form as for the ML-III equation.}
Here
\begin{eqnarray}
B_1&=&w\upsigma_3+2i\upepsilon_2\upsigma_3A_{0y}\label{4.30},\\
A_0&=&\begin{pmatrix} 0&q\\-r& 0\end{pmatrix}\label{4.31},\\
B_0&=&-\frac{i}{2}v\upsigma_3+\begin{pmatrix} 0&-\upepsilon_2q_{xy}+iwq\\\upepsilon_2r_{xy}-iwr& 0\end{pmatrix}\label{4.32},\\
B_{-1}&=&\begin{pmatrix} \upeta&-p\\-k& -\upeta\end{pmatrix}\label{4.33}. 
\end{eqnarray}

Now, we assume that $r=\updelta q^{*}, \quad k=\updelta p^{*}$, where $\updelta=\pm1$. Then, the system (\ref{4.19})--(\ref{4.25}) takes \linebreak the form
 \begin{eqnarray}
iq_{t}+i\upepsilon_2q_{xxy}-vq+(wq)_x-2ip&=&0 \label{4.34},\\
v_{x}-2i\upepsilon_2\updelta(q^{*}_{xy}q-q^{*}q_{xy})&=&0\label{4.35},\\
w_{x}-2i\upepsilon_2\updelta(|q|^2)_y&=&0\label{4.36},\\
p_{x}-2i\upomega p -2\upeta q&=&0\label{4.37},\\
\upeta_{x}+\updelta(q^{*} p +p^{*} q)&=&0\label{4.38}.
\end{eqnarray}
The above set of Equations (4.39)--(4.43) is the reduced form of the (2 + 1)-dimensional cmKdVMB equation. It admits the following integrable reduction, if $\upepsilon_2-1=p=\upeta=0$:
 \begin{eqnarray}
iq_{t}+iq_{xxy}-vq+(wq)_x&=&0 \label{4.39},\\
v_{x}-2i\updelta(q^{*}_{xy}q-q^{*}q_{xy})&=&0\label{4.40},\\
w_{x}-2i\updelta(|q|^2)_y&=&0\label{4.41}.
\end{eqnarray}
It is the usual (2 + 1)-dimensional cmKdV equation. 

 In (1 + 1) dimensions, that is if $y=x$, the cmKdVMBE (\ref{4.34})--(\ref{4.38}) reduces to the \linebreak (1 + 1)-dimensional cmKdVHMBE, which has the form (see, e.g., \cite{C1,C2})
 \begin{eqnarray}
q_{t}+\upepsilon_2(q_{xxx}+6\updelta|q|^2q_x)-2p&=&0 \label{4.42},\\
p_{x}-2i\upomega p -2\upeta q&=&0\label{4.43},\\
\upeta_{x}+\updelta(q^{*} p +p^{*} q)&=&0\label{4.44}.
 \end{eqnarray}
 Its Lax representation reads as
 \begin{eqnarray}
\Psi_{x}&=&A\Psi\label{4.45},\\
\Psi_{t}&=&(4\upepsilon_2\uplambda^2A+B)\Psi\label{4.46}, 
\end{eqnarray} 
where
 \begin{eqnarray}
A&=&-i\uplambda \upsigma_3+A_0\label{4.47},\\
B&=&\uplambda B_1+B_0+\frac{i}{\uplambda+\upomega}B_{-1}\label{4.48}.
\end{eqnarray} 
here
\begin{eqnarray}
B_1&=&2i\upepsilon_2\updelta|q|^2\upsigma_3+2i\upepsilon_2\upsigma_3A_{0y}\label{4.49},\\
A_0&=&\begin{pmatrix} 0&q\\-r& 0\end{pmatrix}\label{4.50},\\
B_0&=&\upepsilon_2\updelta(q^{*}_{x}q-q^{*}q_{x})\upsigma_3+B_{01}\label{4.51}, \\
B_{01}&=&\begin{pmatrix} 0&-\upepsilon_2q_{xx}-2\upepsilon_2\updelta|q|^2q\\\upepsilon_2r_{xx}+2\upepsilon_2\updelta|q|^2r& 0\end{pmatrix}\label{4.52},\\
B_{-1}&=&\begin{pmatrix} \upeta&-p\\-k& -\upeta\end{pmatrix}\label{4.53}. 
\end{eqnarray}
 Note that the (1 + 1)-dimensional cmKdVMBE (\ref{4.42})--(\ref{4.44}) itself admits the following \linebreak integrable reductions.
 
\vspace{6pt}
 (i) The (1 + 1)-dimensional complex mKdV equation is obtained when $\upepsilon_2-1=p=\upeta=0$:
 \begin{eqnarray}
q_{t}+q_{xxx}+6\updelta|q|^2q_x=0 \label{4.54}.\
 \end{eqnarray}

 (ii) The following (1 + 1)-dimensional equation arises when $\upepsilon_2=0$:
 \begin{eqnarray}
q_{t}-2p&=&0 \label{4.55},\\
p_{x}-2i\upomega p -2\upeta q&=&0\label{4.56},\\
\upeta_{x}+\updelta(q^{*} p +p^{*} q)&=&0\label{4.57}.
 \end{eqnarray}
 or:
 \begin{eqnarray}
\frac{1}{2}q_{xt}-i\upomega q_t -2\upeta q&=&0\label{4.58},\\
2\upeta_{x}+\updelta(|q|^{2})_t&=&0\label{4.59}.
 \end{eqnarray}
 
 (iii) On the other hand, the following (1 + 1)-dimensional equation results for $\updelta=0$:
 \begin{eqnarray}
q_{t}+\upepsilon_2q_{xxx}-2p&=&0 \label{4.60},\\
p_{x}-2i\upomega p -2\upeta_0 q&=&0\label{4.61},
 \end{eqnarray}
 where $\upeta_0$ is a constant. Again, we note that all of these reductions are integrable in the Lax sense. The corresponding Lax representations can be obtained from the Lax representation (\ref{4.45}) and (\ref{4.46}) as the associated reductions.


\section {The Myrzakulov--Lakshmanan IV Equation}

Our third new integrable spin system is the Myrzakulov--Lakshmanan IV equation (ML-IV equation), which is a higher order spin evolution equation, 
\begin{eqnarray}
iS_{t}+2\upepsilon_1Z_x+i\upepsilon_2(S_{xy}+[S_x,Z])_{x}+(wS)_{x}+\frac{1}{\upomega}[S, W]&=&0\label{5.1},\\
u_x-\frac{i}{4}tr(S\times[S_x,S_y])&=&0\label{5.2},\\
w_x-\frac{i}{4}\upepsilon_2[tr(S_x^2)]_y&=&0\label{5.3},\\
 iW_{x}+\upomega [S, W]&=&0\label{5.4},
\end{eqnarray} 
 where, again, $Z$ is defined by Equation (4.5). 
 This equation is also integrable in the Lax pair sense. Below, we present some salient features on the ML-IV equation.
\subsection{Lax Representation} 

First let us present the corresponding Lax representation of the ML-IV Equations (\ref{5.1})--(\ref{5.4}). It has the form
 \begin{eqnarray}
\Phi_{x}&=&U\Phi\label{5.6},\\
\Phi_{t}&=&(2\upepsilon_1\uplambda+4\upepsilon_2\uplambda^2)\Phi_y+V\Phi\label{5.7} 
\end{eqnarray} 
with
 \begin{eqnarray}
U&=&-i\uplambda S\label{5.8},\\
V&=&(2\upepsilon_1\uplambda+4\upepsilon_2\uplambda^2)Z+\uplambda V_{1}+\frac{i}{\uplambda+\upomega}W-\frac{i}{\upomega}W\label{5.9}, 
\end{eqnarray} 
where
\begin{eqnarray}
V_1&=&wS+i\upepsilon_2(S_{xy}+[S_x,Z])\label{5.10},\\
W&=&\begin{pmatrix} W_3&W^{-}\\W^{+}& -W_3\end{pmatrix}\label{5.11}, 
\end{eqnarray} 
so that compatibility condition $\Phi_{xt}=\Phi_{tx}$ gives the LM-IV Equations (\ref{5.1})--(\ref{5.4}). Here, $\uplambda$ evolves as 
\begin{eqnarray}
\uplambda_{t}=(2\upepsilon_1 \uplambda+4\upepsilon_2 \uplambda^2)\uplambda_y. 
 \end{eqnarray} 
 
\subsection{Reductions}

Now, we present some reductions of the LM-IV Equations (\ref{5.1})--(\ref{5.4}).
\subsubsection{Case I: $\upepsilon_1=\upepsilon_2=0$}

Let us put $\upepsilon_1=\upepsilon_2=0$. Then, the LM-IV equation reduces to the form
\begin{eqnarray}
iS_{t}+\frac{1}{\upomega}[S, W]&=&0\label{5.12},\\
 iW_{x}+\upomega [S, W]&=&0\label{5.13}.
\end{eqnarray} 
 It is nothing but the principal chiral equation noted previously, which is integrable.
\subsubsection{Case II: $\upepsilon_1\neq 0, \upepsilon_2=0$}

Next, we consider the case $\upepsilon_1\neq 0, \upepsilon_2=0$. Then, we get
\begin{eqnarray}
iS_{t}+2\upepsilon_1([S, S_{y}]+2iuS)_x+\frac{1}{\upomega}[S, W]&=&0\label{5.14},\\
u_x-\frac{i}{4}tr(S\times[S_x,S_y])&=&0\label{5.15},\\
 iW_{x}+\upomega [S, W]&=&0\label{5.16}.
\end{eqnarray} 
It is the ML-II Equations (3.1)--(3.3).
\subsubsection{Case III: $\upepsilon_1=0, \upepsilon_2\neq 0$}

Our next example is the case $\upepsilon_1=0, \upepsilon_2\neq 0$. In this case, the ML-IV equation takes the form
\begin{eqnarray}
iS_{t}+i\upepsilon_2(S_{xy}+[S_x,Z])_{x}+(wS)_{x}+\frac{1}{\upomega}[S, W]&=&0\label{5.17},\\
u_x-\frac{i}{4}tr(S\times[S_x,S_y])&=&0\label{5.18},\\
w_x-\frac{i}{4}\upepsilon_2[tr(S_x^2)]_y&=&0\label{5.19},\\
 iW_{x}+\upomega [S, W]&=&0\label{5.20}.
\end{eqnarray} 
 It is nothing but the ML-III Equations (\ref{4.1})--(\ref{4.4}).
\subsubsection{Case IV: $W=0$}

Now, we put $W=0$. Then, we have
\begin{eqnarray}
iS_{t}+2\upepsilon_1Z_x+i\upepsilon_2(S_{xy}+[S_x,Z])_{x}+(wS)_{x}&=&0\label{5.21},\\
u_x-\frac{i}{4}tr(S\times[S_x,S_y])&=&0\label{5.22},\\
w_x-\frac{i}{4}\upepsilon_2[tr(S_x^2)]_y&=&0\label{5.23}.
\end{eqnarray} 
 
\subsubsection{Case V: $y=x$} 

The last example is the case $y=x$, that is the (1 + 1)-dimensional case. This case corresponds to \linebreak the equation
\begin{eqnarray}
iS_{t}+\frac{1}{2}\upepsilon_1[S,S_{xx}]+i\upepsilon_2(S_{xx}+6tr(S_x^2)S)_{x}+\frac{1}{\upomega}[S, W]&=&0\label{5.24},\\
 iW_{x}+\upomega [S, W]&=&0\label{5.25}.
\end{eqnarray} 
 It is the Myrzakulov-XCIV equation (M-XCIV equation) (see, e.g., \cite{R13} and \cite{ML2014}).

\subsection{Equivalent Counterpart of the ML-IV Equation}

 The gauge equivalent counterpart of the ML-IV Equation (\ref{5.1})--(\ref{5.4}) has the form
 \begin{eqnarray}
iq_{t}+\upepsilon_1q_{xy}+i\upepsilon_2q_{xxy}-vq+(wq)_x-2ip&=&0 \label{5.26},\\
ir_{t}-\upepsilon_1r_{xy}+i\upepsilon_2r_{xxy}+vr+(wr)_x-2ik&=&0\label{5.27},\\
v_{x}+2\upepsilon_1(rq)_y-2i\upepsilon_2(r_{xy}q-rq_{xy})&=&0\label{5.28},\\
w_{x}-2i\upepsilon_2(rq)_y&=&0\label{5.29},\\
p_{x}-2i\upomega p -2\upeta q&=&0\label{5.30},\\
k_x+2i\upomega k-2\upeta r&=&0\label{5.31},\\
\upeta_{x}+r p +k q&=&0\label{5.32}.
\end{eqnarray}
We designate this set of equations as the (2 + 1)-dimensional Hirota--Maxwell--Bloch equation (HMBE) for the reason that when $y=x$, it gives the (1 + 1)-dimensional Hirota--Maxwell--Bloch equation (HMBE) (see, e.g., \cite{C1,C2}). The set of Equations (\ref{5.26})--(\ref{5.32}) is also integrable as it admits the Lax representation. The corresponding Lax representation reads as
\begin{eqnarray}
\Psi_{x}&=&A\Psi\label{5.33},\\
\Psi_{t}&=&(2\upepsilon_1\uplambda+4\upepsilon_2\uplambda^2)\Psi_y+B\Psi\label{5.34},
\end{eqnarray} 
where
 \begin{eqnarray}
A&=&-i\uplambda \upsigma_3+A_0\label{5.35},\\
B&=&\uplambda B_1+B_0+\frac{i}{\uplambda+\upomega}B_{-1}\label{5.36}.
\end{eqnarray} 
\textcolor{black}{Here, the spectral parameter satisfies the following evolution equation $\uplambda_{t}=(2\upepsilon_1\uplambda+4\upepsilon_2\uplambda^2)\uplambda_{y}$ and}
\begin{eqnarray}
B_1&=&w\upsigma_3+2i\upepsilon_2\upsigma_3A_{0y}\label{5.37},\\
A_0&=&\begin{pmatrix} 0&q\\-r& 0\end{pmatrix}\label{5.38},\\
B_0&=&-\frac{i}{2}v\upsigma_3+\begin{pmatrix} 0&i\upepsilon_1q_y-\upepsilon_2q_{xy}+iwq\\i\upepsilon_1r_y+\upepsilon_2r_{xy}-iwr& 0\end{pmatrix}\label{5.39},\\
B_{-1}&=&\begin{pmatrix} \upeta&-p\\-k& -\upeta\end{pmatrix}\label{5.40}. 
\end{eqnarray}
Now, we assume that $r=\updelta q^{*}, \quad k=\updelta p^{*}$, where $\updelta=\pm1$. Then, the system (\ref{5.26})--(\ref{5.32}) takes the form
 \begin{eqnarray}
iq_{t}+\upepsilon_1q_{xy}+i\upepsilon_2q_{xxy}-vq+(wq)_x-2ip&=&0 \label{5.41},\\
v_{x}+2\upepsilon_1\updelta(|q|^2)_y-2i\upepsilon_2\updelta(q^{*}_{xy}q-q^{*}q_{xy})&=&0\label{5.42},\\
w_{x}-2i\upepsilon_2\updelta(|q|^2)_y&=&0\label{5.43},\\
p_{x}-2i\upomega p -2\upeta q&=&0\label{5.44},\\
\upeta_{x}+\updelta(q^{*} p +p^{*} q)&=&0\label{5.45}.
 \end{eqnarray}
 It is the (2 + 1)-dimensional HMBE (compare with the HMBE from \cite{C1,C2}). This equation admits the following integrable reductions.
 
 (i) For the case $\upepsilon_1-1=\upepsilon_2=p=\upeta=0$, we get
 \begin{eqnarray}
iq_{t}+q_{xy}-vq&=&0 \label{5.46},\\
v_{x}+2\updelta(|q|^2)_y&=&0\label{5.47}
 \end{eqnarray}
 which is the well-known (2 + 1)-dimensional nonlinear Schr\"odinger equation \cite{myrzakulov-391}.
 
 (ii) The (2 + 1)-dimensional complex mKdV equation is obtained for the choice $\upepsilon_1=\upepsilon_2-1=p=\upeta=0$:
 \begin{eqnarray}
iq_{t}+i\upepsilon_2q_{xxy}-vq+(wq)_x&=&0 \label{5.48},\\
v_{x}-2i\upepsilon_2\updelta(q^{*}_{xy}q-q^{*}q_{xy})&=&0\label{5.49},\\
w_{x}-2i\upepsilon_2\updelta(|q|^2)_y&=&0\label{5.50}.
 \end{eqnarray}
 
 (iii) The (2 + 1)-dimensional Schr\"odinger--Maxwell--Bloch equation results when $\upepsilon_1-1=\upepsilon_2=0$:
 \begin{eqnarray}
iq_{t}+q_{xy}-vq+(wq)_x-2ip&=&0 \label{5.51},\\
v_{x}+2\updelta(|q|^2)_y&=&0\label{5.52},\\
p_{x}-2i\upomega p -2\upeta q&=&0\label{5.53},\\
\upeta_{x}+\updelta(q^{*} p +p^{*} q)&=&0\label{5.54}.
 \end{eqnarray}
 
 (iv) \scalebox{.98}[1.0]{The (2 + 1)-dimensional complex mKdV-Maxwell--Bloch equation is obtained for $\upepsilon_1=\upepsilon_2-1=0$} (see, for example, Equations (4.39)--(4.43)).

 (v) The following (2 + 1)-dimensional equation is obtained for $\updelta=0$:
 \begin{eqnarray}
iq_{t}+\upepsilon_1q_{xy}+i\upepsilon_2q_{xxy}-2ip&=&0 \label{5.65},\\
p_{x}-2i\upomega p -2\upeta_0 q&=&0\label{5.66},
 \end{eqnarray}
 where $\upeta_0=0$. Again, we note that all of these reductions admit the Lax representations and, in this~sense, are integrable. The corresponding Lax representations are obtained from the Lax representation (\ref{5.33}) and (\ref{5.34}) as the corresponding reductions.
In (1 + 1) dimensions, that is if $y=x$, this system reduces to the (1 + 1)-dimensional HMBE, which has the form (see, e.g., \cite{C1,C2})
 \begin{eqnarray}
iq_{t}+\upepsilon_1(q_{xx}+2\updelta|q|^2q)+i\upepsilon_2(q_{xxx}+6\updelta|q|^2q_x)-2ip&=&0 \label{5.67},\\
p_{x}-2i\upomega p -2\upeta q&=&0\label{5.68},\\
\upeta_{x}+\updelta(q^{*} p +p^{*} q)&=&0\label{5.69}.
 \end{eqnarray}
 Its Lax representation reads as
 \begin{eqnarray}
\Psi_{x}&=&A\Psi\label{5.70},\\
\Psi_{t}&=&((2\upepsilon_1\uplambda+4\upepsilon_2\uplambda^2)A+B)\Psi\label{5.71}, 
\end{eqnarray} 
where
 \begin{eqnarray}
A&=&-i\uplambda \upsigma_3+A_0\label{5.72},\\
B&=&\uplambda B_1+B_0+\frac{i}{\uplambda+\upomega}B_{-1}\label{5.73}.
\end{eqnarray} 
here
\begin{eqnarray}
B_1&=&2i\upepsilon_2\updelta|q|^2\upsigma_3+2i\upepsilon_2\upsigma_3A_{0y}\label{5.74},\\
A_0&=&\begin{pmatrix} 0&q\\-r& 0\end{pmatrix}\label{5.75},\\
B_0&=&(i\upepsilon_1\updelta|q|^2+\upepsilon_2\updelta(q^{*}_{x}q-q^{*}q_{x}))\upsigma_3+B_{01} \label{5.76},\\
B_{01}&=&\begin{pmatrix} 0&i\upepsilon_1q_x-\upepsilon_2q_{xx}-2\upepsilon_2\updelta|q|^2q\\i\upepsilon_1r_x+\upepsilon_2r_{xx}+2\upepsilon_2\updelta|q|^2r& 0\end{pmatrix}\label{5.77},\\
B_{-1}&=&\begin{pmatrix} \upeta&-p\\-k& -\upeta\end{pmatrix}\label{5.78}.
\end{eqnarray}
 Note that the (1 + 1)-dimensional HMBE (\ref{5.67})--(\ref{5.69}) admits the following integrable reductions.
 
\vspace{6pt}
 (i) The NSLE for $\upepsilon_1-1=\upepsilon_2=p=\upeta=0$:
 \begin{eqnarray}
iq_{t}+q_{xx}+2\updelta|q|^2q=0\label{5.79}.
 \end{eqnarray}
 
 (ii) The (1 + 1)-dimensional complex mKdV equation for $\upepsilon_1=\upepsilon_2-1=p=\upeta=0$:
 \begin{eqnarray}
q_{t}+q_{xxx}+6\updelta|q|^2q_x=0 \label{5.80}.
 \end{eqnarray}
 
 (iii) The (1 + 1)-dimensional Schr\"odinger--Maxwell--Bloch equation for $\upepsilon_1-1=\upepsilon_2=0$:
 \begin{eqnarray}
iq_{t}+q_{xx}+2\updelta|q|^2q-2ip&=&0 \label{5.81},\\
p_{x}-2i\upomega p -2\upeta q&=&0\label{5.82},\\
\upeta_{x}+\updelta(q^{*} p +p^{*} q)&=&0\label{5.83}.
\end{eqnarray}
 
 (iv) The (1 + 1)-dimensional complex mKdV-Maxwell--Bloch equation for $\upepsilon_1=\upepsilon_2-1=0$ (see Equations (4.47)--(4.49)).
 
 (v) The following (1 + 1)-dimensional equation is obtained for $\upepsilon_1=\upepsilon_2=0$:
 \begin{eqnarray}
 q_{t}-2p&=&0 \label{5.87},\\
p_{x}-2i\upomega p -2\upeta q&=&0\label{5.88},\\
\upeta_{x}+\updelta(q^{*} p +p^{*} q)&=&0\label{5.89}.
 \end{eqnarray}
It can be rewritten in the following form 
\begin{eqnarray}
\frac{1}{2}q_{xt}-i\upomega q_t -2\upeta q&=&0\label{5.90},\\
2\upeta_{x}+\updelta(|q|^{2})_t&=&0\label{5.91}.
 \end{eqnarray}
 
 (vi) The following (1 + 1)-dimensional equation is obtained for $\updelta=0$:
 \begin{eqnarray}
iq_{t}+\upepsilon_1q_{xx}+i\upepsilon_2q_{xxx}-2ip&=&0 \label{5.92},\\
p_{x}-2i\upomega p -2\upeta_0 q&=&0\label{5.93},
 \end{eqnarray}
 where $\upeta_0=0$. Again, we note that all of these reductions are integrable, as they admit Lax~representations. The corresponding Lax representations can be obtained from the Lax \linebreak representation (\ref{5.70}) and (\ref{5.71}) as appropriate reductions.
 
\section{Conclusion}

Spin systems are fascinating nonlinear dynamical systems. In particular, integrable spin systems have much relevance in applied ferromagnetism and nanomagnetism. More interestingly, integrable spin systems have a close connection to the nonlinear Schr\"odinger family of equations. In this paper, we have introduced three specific cases of (2 + 1)-dimensional integrable spin systems, which we designated as the Myrzakulov--Lakshmanan II, III and IV equations, where additional scalar potentials or vector potentials interact in specific ways with the spin fields. Through appropriate gauge or geometric equivalence, we have identified the three equivalent \linebreak (2 + 1)-dimensional nonlinear Schr\"odinger family of equations along with their Lax pairs. These equations, in turn, encompass a large class of the interesting (2 + 1)-dimensional family of NLS
 equations. Regarding both the (2 + 1)-dimensional spin and NLS family of equations, an extremely interesting question is to investigate what the the physical applications of these new equations are, for example in nonlinear optics (for their (1 + 1)-dimensional analogues, see, e.g., \cite{C2,Beggs,ML2014,Porsezian}). It will be also interesting to investigate the other associated integrability properties, like the infinite number of conservation laws, the involutive integrals of motion, soliton solutions, exponentially-localized dromion solutions, \emph{etc.} In particular, it is interesting to investigate the relation between the above presented integrable spin systems with self-consistent potentials and the geometry of curves and surfaces (see, e.g.,  \cite{Es, M0, MK7,myrzakulov-9535,myrzakulov-715,myrzakulov-83,myrzakulov-233,myrzakulov-378,myrzakulov-1576,myrzakulov-543,myrzakulov-535,myrzakulov-248,myrzakulov-314,Zhao,Yan,Zhen-Huan,myrzakulov-1397, Gabitov}). Work is in progress along these lines, and the results will be \linebreak reported separately.

 
\section*{Acknowledgments}\addcontentsline{toc}{section}{Acknowledgments}
The work of Muthusamy Lakshmanan is supported by DST--IRHPA research project. The work of  Muthusamy Lakshmanan is also supported by a DAE  Raja Ramanna Fellowship.
\section*{Author Contributions}\addcontentsline{toc}{section}{Author Contributions}
All authors have equally contributed to every aspect of the paper.

\section*{Conflicts of Interest}\addcontentsline{toc}{section}{Conflicts of Interest}
The authors declare no conflict of interest. 

\section*{Appendix A: Gauge equivalence}\addcontentsline{toc}{section}{Appendix A: Gauge equivalence}
In the previous sections, we presented some new integrable spin systems in 2 + 1 
 dimensions. Furthermore, we presented their equivalent counterparts in terms of the NLS family of equations. Here, in this Appendix, we want to demonstrate that between the former and latter systems, a gauge equivalence takes place. As an example, we consider the ML-IV equation, as the other two systems are its particular cases. Let $\Phi(t,x, \uplambda)$ be the solution of System (\ref{5.6}) and (\ref{5.7}) and $\Psi(t,x, \uplambda)$ be the solution of System (\ref{5.33}) and (\ref{5.34}). Then, it is not difficult to verify that these functions are related by the tranformation
$$
\Psi=g\Phi\label{7.1},\eqno{(A1)}
$$
 where $g(t,x)=\Psi(t,x,0)$.
 This means that between the ML-IV Equations (\ref{5.1})--(\ref{5.4}) and the nonlinear Schr\"odinger-type System (\ref{5.26})--(\ref{5.32}), the gauge equivalence takes place. Hence, in particular, some relations follow between $W$ and its mirrors $p,k$ and $\upeta$. The vector potential ${\bf W}=(W_1, W_2, W_3)$ or its matrix form $W$ obeys the condition
$$
W^2=(W_3^2+W^{+}W^{-})I=C(t)I\label{7.2}.\eqno{(A2)}
$$
where in the standard spin cases, we set $C(t)=constant$. On the other hand, from the \linebreak Equations (\ref{5.30})--(\ref{5.32}), it follows that
$$
\upeta^2+pk=C(t)\label{7.3}\eqno{(A3)}
$$
 so that we have
 $$
W^2=(\upeta^2+pk)I\label{7.4}\eqno{(A4)}
$$
 or
$$
W_3^2+W^{+}W^{-}=\upeta^2+pk=C(t)\label{7.5}.\eqno{(A5)}
$$
 
\section*{Appendix B: Nonisospectral problem}\addcontentsline{toc}{section}{Appendix B: Nonisospectral problem}

Finally, we would like to note that the nonlinear equations considered in this paper correspond to the nonisospectral problem. In fact, for example, for the more general ML-IV equation, the spectral parameter obeys the equation
$$
\uplambda_t=(2\upepsilon_1\uplambda+4\upepsilon_2\uplambda^2)\uplambda_y\label{8.1}.\eqno{(B1)}
$$
 Let us in more detail consider the case when $\upepsilon_1=1, \quad \upepsilon_2=0$. Then, Equation (B1) takes the form
 $$
\uplambda_t=2\uplambda\uplambda_y \label{8.2}.\eqno{(B2)}
$$ Here, some comments are in order.
 
\vspace{6pt} 
 (i) This last equation has, for example, the following particular solution:
 $$
\uplambda=(\upbeta_1+\upbeta_2y)(\upbeta_3-2\upbeta_2t)^{-1}\label{8.3},\eqno{(B3)}
 $$ where $\upbeta_j$ are in general some complex constants. 
 
 (ii) In the more general case, Equation (B2) admits the solution 
$$
\uplambda=f(y-\frac{1}{2}5t\uplambda)\label{8.4},\eqno{(B4)}
$$ where $f$ is some arbitrary function. Such types of solutions are breaking waves \cite{Brunelli}. 
 
 (iii) Equation (B2) is in fact the dispersionless KdV equation, where the KdV equation itself has \linebreak the form 

$$
\uplambda_t=2\uplambda\uplambda_y+\uplambda_{yyy}. \eqno{(B5)}
$$ 
 
 (iv) Equation (B2) is sometimes called the Riemann equation \cite{Whitham}. 
 
 (v) Equation (B2) is integrable. The corresponding Lax representation is given by
 (see, e.g., \cite{Brunelli})
$$
L_t=\{L, M\}\label{8.5},\eqno{(B6)}
$$
 where
$$
L=p^2+\uplambda, \quad M=p^3+2\uplambda p\label{8.6}.\eqno{(B7)}
$$
 Here, the bracket $\{,\}$ has the form
$$
\{A,B\}=\frac{\partial A}{\partial p}\frac{\partial B}{\partial x}-\frac{\partial B}{\partial p}\frac{\partial A}{\partial x}\label{8.7}.\eqno{(B8)}
$$


\begin{thebibliography}{the 99}

\bibitem{royal}
Lakshmanan, M. The Fascinating World of Landau-Lifshitz-Gilbert Equation: An Overview. \emph{Phil. Trans. R. Soc. A} \textbf{2011}, \emph{369}, 1280-1300.

\bibitem{hillebrands}

Hillebrands, B.; Ounadjela, K. \textit{Spin Dynamics in Confined Magnetic Structures}; Springer-Verlag: Berlin, Germany, 2002; Volumes I and II.

\bibitem{bertotti}
Bertotti, G.; Mayergoyz, I.; Serpico, C. \textit{Nonlinear Magnetization Dynamics in Nanosystems}; Elsevier: Amsterdam, The Netherlands, 2009.

\bibitem{stiles}
Stiles, M.D.; Miltat, J. Spin Transfer Torque and Dynamics. \emph{Top. Appl. Phys.} {\bf 2006}, \emph{101}, 225-308.

\bibitem{slonczewski}
Slonczewski, J.C. Current-driven excitation of magnetic multilayers.
 \emph{J. Magn. Magn. Mater.} {\bf 1996}, \emph{159}, L261-L268.

\bibitem{lakshmanan77}
Lakshmanan, M. Continuum spin system as an exactly solvable dynamical system. \emph{Phys. Lett. A.} {\bf 1977}, \emph{61}, 53-54.

\bibitem{takhtajan}
Takhtajan, L.A. Integration of the continuous Heisenberg spin chain through the
inverse scattering method. \emph{Phys. Lett. A.} {\bf 1977}, \emph{64}, 235-238.

\bibitem{senthilkumar}
Senthilkumar, C.; Lakshmanan, M.; Grammaticos, B.; Ramani, A.Nonintegrability
of (2+1) dimensional continuum isotropic Heisenberg spin system: Painleve analysis.  \emph{Phys. Lett. A.} {\bf 2006}, \emph{356}, 339-345.

\bibitem{ishimori}
Ishimori, Y. Multi-vortex solutions of a two-dimensional nonlinear wave equation.  \emph{Prog. Theor. Phys.} {\bf 1984}, \emph{72}, 33-37.

\bibitem{myrzakulov-391}	 Myrzakulov, R.; Vijayalakshmi, S.; Nugmanova, G.; Lakshmanan, M. A (2+ 1)-dimensional integrable spin model: Geometrical and gauge equivalent counterpart, solitons and localized coherent structures. \emph{Phys. Lett. A.} {\bf 1997}, \emph{233}, 391-396. 
	
\bibitem{ablowitz}
Ablowitz, M.J.; Clarkson, P.A. \textit{Solitons, Nonlinear Evolution Equations Inverse Scattering}; Cambridge University Press: New York, NY, USA, 1991.

\bibitem{myrzakulov-2122}	 Myrzakulov, R.; Vijayalakshmi, S.; Syzdykova, R.; Lakshmanan, M. On the simplest (2+1)-- dimensional integrable spin systems and their equivalent nonlinear Schrodinger equations.  \emph{J. Math. Phys.} {\bf 1998}, \emph{39}, 2122-2140. 	

\bibitem{myrzakulov-3765} Lakshmanan, 	M.; Myrzakulov, R.; Vijayalakshmi, S.; Danlybaeva, A. Motion of curves and surfaces and nonlinear evolution equations in (2+1) dimensions. \emph{J. Math. Phys.} {\bf 1998}, \emph{39}, 3765-3771. 	

\bibitem{K2} Konopelchenko, B.G.; Matkarimov, B.T. {Inverse spectral transform for the Ishimori equation: I. Initial value problem}. \emph{J. Math. Phys.} {\bf 1990}, \emph{31}, 2737-2746. 
\bibitem{Fokas} Fokas, A.S.; Santini, P.M. Dromions and a boundary value problem for the Davey-Stewartson 1 equation.  \emph{Phys. D: Nonlinear Phenom.} {\bf 1990}, \emph{44}, 99-130.
\bibitem{K1} Konopelchenko, B.G. \textit{ Solitons in Multidimensions: Inverse Spectral Transform}; World Scientific: Singapore, 1993.

\bibitem{Chen} Chen, C.; Zhou, Z.-X. Darboux Transformation and Exact Solutions of the Myrzakulov--I Equation. \emph{Chin. Phys. Lett.} {\bf 2009}, \emph{26}, 080504.

\bibitem{Hai} Chen, H.; Zhou, Z.-X. Darboux Transformation with a Double Spectral Parameter for the Myrzakulov--I Equation.  \emph{ Chin. Phys. Lett.} {\bf 2014}, \emph{31}, 120504. 

\bibitem{Zakharov} Zakharov, V.E. The Inverse Scattering Method. In \textit{Solitons}; Bullough, R.K., Caudrey, P.J., Eds.; Springer: Berlin, Germany, 1980.
\bibitem{Strachan1} Strachan, I.A.B. Some integrable hierarchies in (2+1) dimensions and their twistor description.  \emph{J. Math. Phys.} {\bf 1993}, \emph{34}, 243-259.
\bibitem{Strachan2} Strachan, I.A.B. Wave solutions of a (2+1)--dimensional generalization of the nonlinear Schrodinger equation.  \emph{Inverse  Problems.} {\bf 1992}, \emph{8}, L21.

\bibitem{Calogero1} Calogero, F.; Degasperis, A. Extension of the Spectral Transform Method for Solving Nonlinear Evolution Equations. \emph{Lett. Nuovo. Cimento.}. 
{\bf 1978}, \emph{22}, 131--137;  
\bibitem{Calogero2} Calogero, F.; Degasperis, A. Extension of the Spectral Transform Method for Solving Nonlinear Evolution Equations. II. \emph{Lett. Nuovo. Cimento.} {\bf 1978}, \emph{22}, 263--269.
\bibitem{LB} Lakshmanan, M.; Bullough, R.K. Geometry of generalised nonlinear Schrodinger and Heisenberg ferromagnetic spin equations with linearly x-dependent coefficients.  \emph{Phys. Lett. A.} {\bf 1980}, \emph{80}, 287--292.
\bibitem{Bala1} Balakrishnan, R. Inverse spectral transform analysis of a nonlinear Schrodinger equation with x--dependent coefficients.  \emph{Physica D: Nonlinear Phenomena.} {\bf 1985}, \emph{16}, 405--413.
\bibitem{L31} Lakshmanan, M.; Ganesan, S. Geometrical and gauge equivalence of the generalized hirota, Heisenberg and wkis equations with linear inhomogeneities. \emph{Phys. A: Stat. Mech. Appl.} {\bf 1985}, \emph{132}, 117--142.

\bibitem{Bala2} Balakrishnan, R. Dynamics of a generalised classical Heisenberg chain. \emph{Phys. Lett. A.} {\bf 1982}, \emph{92}, 243--246.

\bibitem{Bala3} Blumenfeld, R.; Balakrishnan, R. Exact multi--twist solutions to the Belavin--Polyakov equation and applications to magnetic systems. \emph{J. Phys. A: Math. Gen.} {\bf 2000}, \emph{33}, 2459--2468.

\bibitem{Es} Esmakhanova, K.R.; Nugmanova, G.N.; Zhao, W.-Z.; Wu, K. Integrable Inhomogeneous Lakshmanan-Myrzakulov Equation. arXiv:nlin/0604034. 
\bibitem{R13} Zhunussova, Z.K.; Yesmakhanova, K.R.; Tungushbaeva, D.I.; Mamyrbekova, G.K.;  Nugmanova, G.N.; Myrzakulov, R. Integrable Heisenberg Ferromagnet Equations with self-consistent potentials.  arXiv:1301.1649.


\bibitem{Calogero3} Calogero, F. A Method to Generate Solvable Nonlinear Evolution Equations.  \emph{Lett. Nuovo Cimento} {\bf 1975}, \emph{14}, 443--448.
\bibitem{Calogero4} Calogero, F.; Degasperis, A. Solution by the spectral transform method of a nonlinear evolution equation including as a special case the cylindrical KdV equation. \emph{ Lett. Nuovo Cimento.} {\bf 1978}, \emph{23}, 150--154.
\bibitem{Sa} Sakhnovich, A. Nonisospectral integrable nonlinear equations with external potentials and their GBDT solutions.  \emph{J. Phys. A: Math. Theor.} {\bf 2008}, \emph{41}, 155204.

\bibitem{C1} Li, C.; He, J.; Porsezian, K. Rogue waves of the Hirota and the Maxwell-Bloch equations.   \emph{Phys. Rev. E.} {\bf 2013}, \emph{87}, 012913.

\bibitem{C2} Li, C.; He, J. Darboux transformation and positons of the inhomogeneous Hirota and the Maxwell-Bloch equation. \emph{Sci. China Phys. Mech. Astron.} {\bf 2014}, \emph{57}, 898.

\bibitem{Beggs} Beggs, E.J. J. Beggs, Solitons in the chiral equations. \emph{ Commun. Math. Phys.} {\bf 1990}, \emph{128}, 131--139.


\bibitem{ML2014} Myrzakulov, R.; Mamyrbekova, G.K.; Nugmanova, G.N.; Yesmakhanova, K.R.; Lakshmanan, M. Integrable motion of curves in self-consistent potentials: Relation to spin systems and soliton equations.  \emph{Phys. Lett. A.} {\bf 2014}, \emph{378}, 2118--2123.

\bibitem{Porsezian}  Porsezian, K.; Nakkeeran, K. Optical Soliton Propagation in an Erbium Doped Nonlinear Light Guide with Higher Order Dispersion. \emph{Phys. Rev. Lett.} {\bf 1995}, \emph{74}, 2941.
\bibitem{M0}Myrzakulov, R. \emph{On Some Integrable and Nonintegrable Soliton Equations of Magnets I-IV}; HEPI:Alma-Ata, Kazakhstan,  1987.

\bibitem{MK7} Myrzakulov, R.; Danlybaeva, A.K.; Nugmanova, G.N.	 Geometry and multidimensional soliton equations. \emph{Theor. Math. Phys. }
{\bf 1999}, \emph{118}, 347--356.


\bibitem{myrzakulov-9535}	 Myrzakulov, R.; Nugmanova, G.; Syzdykova, R.Gauge equivalence between (2+1)-dimensional continuous Heisenberg ferromagnetic models and nonlinear Schrodinger-type equations \emph{J. Phys. A: Math. Theor.} {\bf 1998}, \emph{31}, 9535. 	

\bibitem{myrzakulov-715} 	 Myrzakulov, R.; Daniel, M.; Amuda, R.  Nonlinear spin-phonon excitations in an inhomogeneous compressible biquadratic Heisenberg spin chain. Physica A234, 715-724 (1997)  \emph{Physica A.} {\bf 1997}, \emph{234}, 715-724. 
	
\bibitem{myrzakulov-83}	Myrzakulov, R.; Makhankov, V.G.; Pashaev, O. Gauge equivalence SUSY and classical solutions of OSPU(1,1/1)-Heisenberg model and nonlinear Schrodinger equation. \emph{Lett. Math. Phys.} {\bf 1989}, \emph{16}, 83-92.

\bibitem{myrzakulov-233} Myrzakulov, R.; Makhankov, V.G.; Makhankov, A.General Coherent States and the Continuous Heisenberg XYZ Model with One-Ion Anizotropy. \emph{Phys. Scr.} {\bf 1987}, \emph{35}, 233-237.

	
\bibitem{myrzakulov-378}	 Myrzakulov, R.; Pashaev, O.; Kholmurodov, K. Particle-line excitations in Multicomponent Magnon-Poton System. \emph{Phys. Scr.} {\bf 1986}, \emph{33}, 378-384. 
 
\bibitem{myrzakulov-1576}	Anco, S.C.; Myrzakulov, R. Integrable generalizations of Schrodinger maps and Heisenberg spin models from Hamiltonian flows of curves and surfaces. \emph{J. Geom. Phys.} {\bf 2010}, \emph{60}, 1576--1603.

\bibitem{myrzakulov-543} Myrzakulov, R.; Rahimov, F.K.; Myrzakul, K.; Serikbaev, N.S. {On the geometry of stationary Heisenberg ferromagnets}. In \textit{Non-Linear Waves: Classical and Quantum Aspects}; Kluwer Academic Publishers: Dordrecht, The Netherlands, 2004; pp. 543--549.

\bibitem{myrzakulov-535} Myrzakulov, R.; Serikbaev, N.S.; Myrzakul, K.; Rahimov, F.K. On continuous limits of some generalized compressible Heisenberg spin chains. \emph{J. NATO Sci. Ser. II. Math. Phys. Chem.} {\bf 2004}, \emph{153}, 535-542.

\bibitem{myrzakulov-248} Myrzakulov, R.; Martina, L.; Kozhamkulov, T.A.; Myrzakul, K. {Integrable Heisenberg ferromagnets and soliton geometry of curves and surfaces}. In \textit{Nonlinear Physics: Theory and Experiment. II}; World Scientific: London, UK, 2003; p. 248.

\bibitem{myrzakulov-314} Myrzakulov, R. {Integrability of the Gauss-Codazzi-Mainardi equation in 2 + 1 dimensions}. In \textit{Mathematical Problems of Nonlinear Dynamics}, Proceedings of the International Conference ``Progress in Nonlinear sciences'', Nizhny Novgorod, Russia, 2--6 July 2001; Volume 1, \linebreak pp. 314--319. 



\bibitem{Zhao} Yan, Z.-W.; Chen, M.-R.; Wu, K.; Zhao, W.-Z. (2+1)-Dimensional Integrable Heisenberg Supermagnet Model.  \emph{J. Phys. Soc. Jpn.} {\bf 2012}, \emph{81}, 094006.
\bibitem{Yan} Yan Z.-W.; Chen, M.-R.; Wu, K.; Zhao, W.-Z. Integrable Deformations of the (2+1)-Dimensional Heisenberg Ferromagnetic Model.   \emph{Commun. Theor. Phys.} {\bf 2012}, \emph{58}, 463.



\bibitem{Zhen-Huan} Zhang, Z.-H.; Deng, M.; Zhao, W.-Z.; Wu, K. \textit{On the Integrable Inhomogeneous Myrzakulov-I Equation}; arXiv: nlin/0603069. 


\bibitem{myrzakulov-1397} Martina, L.; Myrzakul, Kur.; Myrzakulov, R.; Soliani, G. 
Deformation of surfaces, integrable systems, and Chern?imons theory.
 \emph{J. Math. Phys.} {\bf 2001}, \emph{42}, 1397.

\bibitem{Gabitov} Burtsev, S.P.; Gabitov, I.R.  Alternative integrable equations of nonlinear optics.  \emph{Phys. Rev. A. } {\bf 1994}, \emph{49}, 2065--2070.

\bibitem{Brunelli} Brunelli, J.C. Dispersionless limit of integrable models.  \emph{Braz. J. Phys.} {\bf 2000}, \emph{30}, 455--468.

\bibitem{Whitham} Whitham, G.B. \textit{Linear and Nonlinear Waves}; John Wiley \& Sons: New York, USA, 1974.
\end{thebibliography}
\end{document}